# Robustness and Overcoming Brittleness of AI-Enabled Legal Micro-Directives: The Role of Autonomous Levels of AI Legal Reasoning

**Dr. Lance B. Eliot**
Chief AI Scientist, Techbruim; Fellow, CodeX: Stanford Center for Legal Informatics
Stanford, California, USA

**Abstract**

Recent research by legal scholars suggests that the law might inevitably be transformed into legal micro-directives consisting of legal rules that are derived from legal standards or that are otherwise produced automatically or via the consequent derivations of legal goals and then propagated via automation for everyday use as readily accessible lawful directives throughout society. This paper examines and extends the legal micro-directives theories in three crucial respects: (1) By indicating that legal micro-directives are likely to be AI-enabled and evolve over time in scope and velocity across the autonomous levels of AI Legal Reasoning, (2) By exploring the tradeoffs between legal standards and legal rules as the imprinters of the micro-directives, and (3) By illuminating a set of brittleness exposures that can undermine legal micro-directives and proffering potential mitigating remedies to seek greater robustness in the instantiation and promulgation of such AI-powered lawful directives.

## 1 Background on Legal Micro-Directives

Recent research by legal scholars advises that the law might inevitably be transformed into legal micro-directives consisting of legal rules that are derived from legal standards or that are otherwise produced automatically or via the consequent derivations of legal goals [8] [9] [28] [32]. These legal micro-directives would be propagated via automation for everyday use as readily accessible, lawful directives, throughout society.

This paper examines and extends the legal micro-directives theories in three crucial respects:

(1) By indicating that legal micro-directives are likely to be AI-enabled and evolve over time in scope and velocity across the autonomous levels of AI Legal Reasoning [20] [22],

(2) By exploring the tradeoffs between legal standards and legal rules as the imprinters of the micro-directives, and

(3) By illuminating a set of brittleness exposures that can undermine legal micro-directives and proffering potential mitigating remedies to seek greater robustness in the instantiation and promulgation of such AI-enabled lawful directives.

In Section 1 of this paper, the topic of legal micro-directives is introduced and addressed. Doing so establishes the groundwork for the subsequent sections. Section 2 introduces the Levels of Autonomy (LoA) of AI Legal Reasoning (AILR), which is instrumental in the discussions undertaken in Section 3. Section 3 provides an indication of the AI-enablement of legal micro-directives, along with exploring the tradeoffs of legal rules versus legal standards, and the brittleness facets thereof. The final section, Section 4, covers additional considerations and recommendations for future research.

This paper then consists of these four sections:

- Section 1: Background on Legal Micro-Directives
- Section 2: Autonomous Levels of AI Legal Reasoning

**1.1 Legal Micro-Directives**

In the research literature about the law, there has been a longstanding philosophical dialogue about the nature of legal standards and legal rules as two keystone elements of the law [2] [25] [26] [28]. The general view is that legal standards are typically indicative of broad strokes about the delineation of lawful behavior while legal rules are more tightly specified. Legal standards are considered open to interpretation and flexible, thus able to encompass a wide variety of acts and activities that would be assessed as lawful, but do so at the potential peril or downside that loose interpretation can lead to unlawful efforts that seemingly were feasible within the scope implied. Legal rules on the other hand are generally considered of a specific nature, being more pinpoint descriptive and reducing interpretive uncertainty about what the law portends [37] [42] [49].

Casey and Niblett [09] indicate this about legal standards and legal rules: "Rules are precise and ex ante in nature. Rules indicate to an individual whether certain behavior will violate or comply with the law. When a rule is enacted, effort must be undertaken by lawmakers to give full and precise content to the law before the individuals act. Standards, on the other hand, are imprecise when they are enacted. The exact content of the law comes after an individual acts, as judges and other adjudicators determine whether the individual's specific behavior in a particular context violates the standard."

Furthermore, with respect to uncertainty, they indicate [09]: "Uncertainty about the content of a law is greater with standards than with simple rules. When regulated by a simple rule, an individual will more likely know whether her behavior is allowed or prohibited. When regulated by a standard, on the other hand, the individual does not know how any particular judge with wide discretion will apply the standard to the facts. She may not know what behavior a judge will consider reasonable."

To readily depict this difference between legal standards and legal rules, a frequently provided example is the use case of automobile driving and laws related to driving that are legal standards versus driving-related legal rules. Per McGinnis and Wasick [32]: "The prototypical example of a rule-based law is a speed limit that holds that a driver must drive at sixty-five miles per hour or less. In contrast, a standard 'requires the judge both to discover the facts of a particular situation and to assess them in terms of the purposes or social values embodied in the standard.' A standards-based speed limit, for example, would hold that a driver must drive at a 'reasonable' speed."

Per the work of Kaplow, as summarized and recounted in [32], legal rules would tend toward presumably greater conformity to the law: "If actors are able to inform themselves as to the consequences of the law beforehand, they are more likely to act in accordance with the law. Under a system of rule-based law, the legal norm is stated before an individual has the opportunity to act, giving them the chance to inform themselves about the law and act accordingly. In a standards-based law, the individual does not know the exact outlines of the law until it is given content by the court. This lack of information would result in less conformity with the law."

With the emergence of computer-based automation that is increasingly becoming ubiquitous throughout society, a relatively new concept or theory has been ventured in the realm of the law that has been coined as the use of legal micro-directives.

Essentially, a legal micro-directive is a legal rule that is made available via automation so that society can be made aware of the legal rule in a real-time and anyplace manner. Casey and Niblett [08] define micro-directives in this way: "Ultimately, law will exist in a catalogue of precisely tailored directives, specifying exactly what is permissible in every unique situation. In this world, when a citizen faces a legal decision, she is informed of exactly how to comply with every relevant law before she acts. The citizen does not have to weigh the reasonableness of her actions nor does she have to search for the content of a law. She follows a simple directive that is optimized for her situation. We call these refined laws 'micro-directives.' These micro-directives will be largely automated." From a lawmaker's perspective, these legal micro-directives are characterized as a new form of law [08]: "The lawmaker's decision between rules and standards will become unnecessary. A new form of law – the micro-directive – will emerge. The micro-directive provides ex ante behavioral prescriptions finely tailored to every possible scenario."

Returning to the example of automobile driving, consider this scenario about the use of legal micro-directives [08]: "To see how the mechanism might work, consider the regulation of traffic speed. In a world of rules and standards, a legislature hoping to optimize safety and travel time could enact a rule (a sixty miles-per-hour speed limit) or a standard ("drive reasonably"). With microdirectives, however, the law looks quite different. The legislature merely states its goal. Machines then design the law as a vast catalog of context-specific rules to optimize that goal. From this catalog, a specific microdirective is selected and communicated to a particular driver (perhaps on a dashboard display) as a precise speed for the specific conditions she faces. For example, a microdirective might provide a speed limit of 51.2 miles per hour for a particular driver with twelve years of experience on a rainy Tuesday at 3:27 p.m. The legislation remains constant, but the microdirective updates as quickly as conditions change."

At first glance, the adoption of legal micro-directives might seem a mere technological advancement and thus otherwise not be especially significant regarding the nature of the law and its ramifications. According to Casey and Niblett [09], they foresee that legal micro-directives could have a potent and enduring impact on law in many vital ways: "First, it will change the broad institutional balance of power in our political and legal system. Second, it may change the development and substantive content of legislative policy. Third, it will transform the practice and training of law. Fourth, it will have moral and ethical consequences for individual citizens, altering their day-to-day decision-making process and changing their relationship with lawmakers and government."

In that impactful sense, it is worthwhile therefore to provide additional attention to the nature and scope of legal micro-directives.

One core element consists of the notion that the legal micro-directives can be readily changed, doing so at a velocity unlike that of today's laws, and furthermore that the legal micro-directives will be better targeted [09]: "The technological changes will allow the law to be more precise, better calibrated, more flexible, more consistent, and less biased."

Likening this to the emergence of autonomous vehicles and self-driving cars, it is said that law will become *self-driving* (borrowing the phrase but not intending to somehow be commingled with autonomous vehicles per se) [09]: "If the state of the world changes, or if the objective of the law is changed, the vast array of micro-directives will instantly update. These laws will be better calibrated, more precise, and more consistent. The law will become, for all intents and purposes, 'self-driving.'"

If one conceives of the legal micro-directives as a variant of the notion of a legal rule, this implies that legal micro-directives will inherit the same advantages of legal rules as stated earlier in comparing the legal standards difficulty versus the benefits of legal rules [32]: "This difficulty is a problem both *ex ante* and *ex post*. *Ex ante*, an actor subject to a law wants to understand what actions the law requires, so that he may avoid liability. Instead of simply looking up a statute, as in the case of a rule, an actor subject to a standard would need to try and collect the relevant case law to determine the outlines of the standard and how it applies to him. *Ex post*, an actor subject to a standard would have to go through the same process to determine the likelihood of success at trial. At trial, the judge has the additional burden of determining the proper application of the standard to the particular facts of the case."

And thus, the benefits of using legal rules [32]**:** "If actors are able to inform themselves as to the consequences of the law beforehand, they are more likely to act in accordance with the law. Under a system of rule-based law, the legal norm is stated before an individual has the opportunity to act, giving them the chance to inform themselves about the law and act accordingly. In a standards-based law, the individual does not know the exact outlines of the law until it is given content by the court. This lack of information would result in less conformity with the law."

Some assert that micro-directives might entirely replace legal standards and legal rules, while others opine that there will still be legal standards, perhaps depicted as legal goals, out of which legal rules will be derived, and then out of which legal micro-directives are specified and promulgated.

In that latter perspective, lawmakers would continue to work at a legal standards or legal goals vantage point, and then be able to automatically have legal rules and legal micro-directives spawned from those legal standards. Those that advocate the contrary posture that all legal standards and legal rules will be eliminated and replaced entirely by micro-directives are apt to argue that there should not be any gaps per se between legal standards and the resultant legal micro-directives, while those that assert the continuing importance of legal standards are apt to suggest that lawmakers would not be readily able to produce law at a legal micro-directive level, and would be more amenable to establishing legal standards or legal goals that would then generate appropriately aligned legal micro-directives.

Open-ended questions about legal micro-directives include whether these legal micro-directives would necessarily be ascertained by humans or might be determined or generated via some form of AI Legal Reasoning (AILR) system or by other automated means [20] [22].

In the work by McGinnis and Wasick [32], they use the phrase "dynamic rules" defined as: "Dynamic rules are rules that automatically change without intervention by the rule giver according to changes in future conditions that the rule itself comprehensively and accurately fixes. As computation increases, it becomes easier to add complex conditions, both because these conditions can be continually monitored and because the application of the new rule can be more readily calculated."

The changes in the legal *dynamic rules* could be tied to data changes [32]: "Dynamic rules are rules that are tied directly to real world empirical data, so that they automatically update as the data to which they are tied changes. Dynamic rules can therefore increase the ability of rules to adapt to continuously changing circumstances rather than await another legislative decision to adapt." An example given to highlight how dynamic rules might function includes the realm of tax brackets [32]: "There are several examples of dynamic rules that are currently in effect, many with very successful results. For instance, the Economic Recovery Tax Act of 1981 indexed tax brackets to inflation. Before this change, taxpayers experienced 'bracket creep' when inflation pushed them into higher tax brackets while their purchasing power remained the same.281 This led to a period during the 1970s when tax brackets had to be frequently changed by Congress in order to keep pace with inflation. By indexing brackets, Congress eliminated the need to revisit tax policy solely due to the inevitable increase of inflation."

Given the tremendous volume of potential legal dynamic rules or legal micro-directives that could end up being produced, it seems unlikely that lawmakers could cope with or manage laws at that granular of a level, and thus would need to be able to continue lawmaking at a higher level [32]: "In theory, rules could also be changed by legislatures or regulatory bodies in response to new information. In practice, however, rules tend to be sticky even in the face of changing circumstances that should modify them. Legislatures tend to be reactive to crises and thus may not update rules continuously as new information becomes available. The legislatures' crowded agendas often make it difficult to find time to update rules."

At the mid-level and granular level of the law, automation potentially consisting of AILR could be the spawning mechanism for legal rules and legal micro-directives, meanwhile, human lawmakers would continue their efforts at higher-levels of lawmaking [09]: "Human policy makers will still play a crucial role. Just as self-driving cars will determine the safest and fastest route to a destination selected by humans, self-driving laws will determine the optimal way to achieve a policy objective chosen by humans. Even though the micro-directives are automated and update in real time, human lawmakers will be required to set the broad objectives of the law."

There are potential downsides to the advent of legal micro-directives, including dystopian possibilities of automation that is restrictive beyond what was intended [08]: "A far more dystopian vision is one where lawmakers turn microdirectives into physical restraints on behavior. Rather than commanding which action should be taken, the individual is restrained from undertaking actions that do not comply with the law. Instead of simply telling the doctor that surgery is not the wisest course of action and that performing surgery will constitute negligence, imagine now that the medical technology required to perform the surgery is automatically switched off, denying the doctor the possibility of performing the surgery."

All told, the upside potential for legal micro-directives in light of the potential adverse consequences of the adoption of legal micro-directives, merits additional research [08]: "The consequences relating to morality, privacy, and autonomy should be addressed before micro-directives arrive."

The next section of this paper introduces the autonomous levels of AI Legal Reasoning, doing so to then aid Section 3 that explores how legal micro-directives might vary across the levels of autonomy. Section 3 also covers more introspection of the legal rules versus legal standards facets and examines how legal micro-directives might be entailed via a macroscopic process flow indication. Also, the potential for brittleness in the use of legal micro-directives is identified, including potential mitigations or remedies that might be leveraged.

## 2 Autonomous Levels of AI Legal Reasoning

In this section, a framework for the autonomous levels of AI Legal Reasoning is summarized and is based on the research described in detail in Eliot [20].

These autonomous levels will be portrayed in a grid that aligns with key elements of autonomy and as matched to AI Legal Reasoning. Providing this context will be useful to the later sections of this paper and will be utilized accordingly.

The autonomous levels of AI Legal Reasoning are as follows:

Level 0: No Automation for AI Legal Reasoning
Level 1: Simple Assistance Automation for AI Legal Reasoning
Level 2: Advanced Assistance Automation for AI Legal Reasoning
Level 3: Semi-Autonomous Automation for AI Legal Reasoning
Level 4: Domain Autonomous for AI Legal Reasoning
Level 5: Fully Autonomous for AI Legal Reasoning
Level 6: Superhuman Autonomous for AI Legal Reasoning

### 2.1 Details of the LoA AILR

See **Figure A-1** for an overview chart showcasing the autonomous levels of AI Legal Reasoning as via columns denoting each of the respective levels.

See **Figure A-2** for an overview chart similar to Figure A-1 which alternatively is indicative of the autonomous levels of AI Legal Reasoning via the rows as depicting the respective levels (this is simply a reformatting of Figure A-1, doing so to aid in illuminating this variant perspective, but does not introduce any new facets or alterations from the contents as already shown in Figure A-1).

#### 2.1.1 Level 0: No Automation for AI Legal Reasoning

Level 0 is considered the no automation level. Legal reasoning is carried out via manual methods and principally occurs via paper-based methods.

This level is allowed some leeway in that the use of say a simple handheld calculator or perhaps the use of a fax machine could be allowed or included within this Level 0, though strictly speaking it could be said that any form whatsoever of automation is to be excluded from this level.

#### 2.1.2 Level 1: Simple Assistance Automation for AI Legal Reasoning

Level 1 consists of simple assistance automation for AI legal reasoning.

Examples of this category encompassing simple automation would include the use of everyday computer-based word processing, the use of everyday computer-based spreadsheets, the access to online legal documents that are stored and retrieved electronically, and so on.

By-and-large, today's use of computers for legal activities is predominantly within Level 1. It is assumed and expected that over time, the pervasiveness of automation will continue to deepen and widen, and eventually lead to legal activities being supported and within Level 2, rather than Level 1.

#### 2.1.3 Level 2: Advanced Assistance Automation for AI Legal Reasoning

Level 2 consists of advanced assistance automation for AI legal reasoning.

Examples of this notion encompassing advanced automation would include the use of query-style

Natural Language Processing (NLP), Machine Learning (ML) for case predictions, and so on.

Gradually, over time, it is expected that computer-based systems for legal activities will increasingly make use of advanced automation. Law industry technology that was once at a Level 1 will likely be refined, upgraded, or expanded to include advanced capabilities, and thus be reclassified into Level 2.

### 2.1.4 Level 3: Semi-Autonomous Automation for AI Legal Reasoning

Level 3 consists of semi-autonomous automation for AI legal reasoning.

Examples of this notion encompassing semi-autonomous automation would include the use of Knowledge-Based Systems (KBS) for legal reasoning, the use of Machine Learning and Deep Learning (ML/DL) for legal reasoning, and so on.

Today, such automation tends to exist in research efforts or prototypes and pilot systems, along with some commercial legal technology that has been infusing these capabilities too.

### 2.1.5 Level 4: Domain Autonomous for AI Legal Reasoning

Level 4 consists of domain autonomous computer-based systems for AI legal reasoning.

This level reuses the conceptual notion of Operational Design Domains (ODDs) as utilized in the autonomous vehicles and self-driving cars levels of autonomy, though in this use case it is being applied to the legal domain [17] [18] [20].

Essentially, this entails any AI legal reasoning capacities that can operate autonomously, entirely so, but that is only able to do so in some limited or constrained legal domain.

### 2.1.6 Level 5: Fully Autonomous for AI Legal Reasoning

Level 5 consists of fully autonomous computer-based systems for AI legal reasoning.

In a sense, Level 5 is the superset of Level 4 in terms of encompassing all possible domains as per however so defined ultimately for Level 4. The only constraint, as it were, consists of the facet that the Level 4 and Level 5 are concerning human intelligence and the capacities thereof. This is an important emphasis due to attempting to distinguish Level 5 from Level 6 (as will be discussed in the next subsection)

It is conceivable that someday there might be a fully autonomous AI legal reasoning capability, one that encompasses all of the law in all foreseeable ways, though this is quite a tall order and remains quite aspirational without a clear cut path of how this might one day be achieved. Nonetheless, it seems to be within the extended realm of possibilities, which is worthwhile to mention in relative terms to Level 6.

### 2.1.7 Level 6: Superhuman Autonomous for AI Legal Reasoning

Level 6 consists of superhuman autonomous computer-based systems for AI legal reasoning.

In a sense, Level 6 is the entirety of Level 5 and adds something beyond that in a manner that is currently ill-defined and perhaps (some would argue) as yet unknowable. The notion is that AI might ultimately exceed human intelligence, rising to become superhuman, and if so, we do not yet have any viable indication of what that superhuman intelligence consists of and nor what kind of thinking it would somehow be able to undertake.

Whether a Level 6 is ever attainable is reliant upon whether superhuman AI is ever attainable, and thus, at this time, this stands as a placeholder for that which might never occur. In any case, having such a placeholder provides a semblance of completeness, doing so without necessarily legitimatizing that superhuman AI is going to be achieved or not. No such claim or dispute is undertaken within this framework.

## 3 Legal Micro-Directives and AI Enablement

In this section, the advent of legal micro-directives is explored in several respects. First, the role of AI as an enabler in the advancement and utility of legal micro-directives is discussed. This discussion includes an indication of the alignment of the evolution of legal micro-directives across the autonomous levels of AI Legal Reasoning. Second, the tradeoffs involved in legal rules versus legal standards are addressed. As part of that analysis, a process flow is proffered on the relationship between legal rules and legal standards as to their deriving the instantiation and promulgation of legal micro-directives. Third, a core set of potential brittleness facets of legal micro-directives is identified. Also, those brittleness concerns are then sought to be overcome via recommended mitigation or remedies, aiming to arrive at a more robust approach to the adoption and use of legal micro-directives.

A series of figures are included in the discussions to aid in illustrating the matters addressed.

### 3.1 Legal Micro-Directives and LoA AILR

As shown in **Figure B-1**, it is useful to align the evolution of legal micro-directives with the autonomous levels of AI Legal Reasoning. AI has the potential to aid in bolstering and strengthening the adoption of legal micro-directives, such that as AI Legal Reasoning improves over time the capabilities can be leveraged toward the creation, derivation, refinement, and application of legal micro-directives.

For each of the levels of autonomy of AI Legal Reasoning, the impacts upon legal micro-directives will be distinctive. A keyword phrasing is used in Figure B-1 to indicate these impacts and consists of:

**LoA AILR – Legal Micro-Directives**

Level 0: *n/a*

Level 1: *Impractical*

Level 2: *Incubatory*

Level 3: *Infancy*

Level 4: *Narrow*

Level 5: *Wide*

Level 6: *Consummate*

In brief, at Level 0, which consists of no automation for AI Legal Reasoning, the applicability to legal micro-directives is considered not applicable ("n/a"), simply due to the by-definition that there is no AI involved at this level. At Level 1, simple assistance automation, the characterization is indicated as "Impractical" since the AI is abundantly unrefined and unable to offer any substantive practical capacity to the legal micro-directive advent. At Level 2, advanced assistance automation, the characterization is indicated as "Incubatory" since the AI at this level can modestly assist in legal micro-directives but is considered quite preliminary in doing so.

At Level 3, the first substantive impact of AI Legal Reasoning comes to work, and this is characterized by the keyword of "Infancy," denoting that the AI is initially being used as a demonstrative enabler for legal micro-directives. Maturing at Level 4, the AI Legal Reasoning is now substantively augmenting the legal micro-directives, yet does so only within particular legal domains, thus this is characterized as being "Narrow" in its impact. Upon Level 5, encompassing all legal domains, the AI Legal Reasoning has now infused across all legal micro-directives and characterized as now being "Wide" in its scope and velocity. Finally, at Level 6, the superhuman AI Legal Reasoning, the advent of micro-directives would be considered "Consummate," though keep in mind that Level 6 is a speculative notion and it is not clear as to what the superhuman capacity would bring forth.

To reiterate and clarify, these depictions are not prescriptive and do not intend to predict what will happen, and instead are a form of taxonomy to depict and describe what might happen and provide an ontological means to understand such phenomena if it should so arise.

### 3.2 Quadrants of Legal Rules Versus Legal Standards

As shown in **Figure B-2**, a four-square set of quadrants is indicative of the tradeoffs of legal rules versus legal standards. Along the rows are the two states or conditions consisting of Legal Rules and the No Legal Rules status. Along the columns are the two states or conditions consisting of Legal Standards and the No Legal Standards status. In combination, there are then four distinct possibilities as represented in the four-square quadrants.

In this case, the quadrants are characterized in this manner:

**Four-square Quadrant of Legal Rules versus Legal Standards**

- Legal Rules: Legal Standards
  *Lawful (via stasis)*
- No Legal Rules: No Legal Standards
  *Lawlessness (a vacuum)*
- Legal Rules: No Legal Standards
  *Infinite Conundrum*
- Legal Standards: No Legal Rules
  *Vagueness Quandary*

In brief, the next subsections consider each of the respective use cases.

### 3.2.1 Legal Rules: Legal Standards – *Lawful (via stasis)*

In this use case, when there are Legal Rules and Legal Standards, the presumption is that this will lead to a state of being lawful, at least concerning guiding as to what is considered lawful behavior. To provide a semblance of a legal direction, the Legal Rules and Legal Standards should be aligned or harmonized, otherwise, there is a potential for confounding of legal micro-directives that attempt to convey what is intrinsically misaligned. To be clear, if there is no overlap or intersection of some Legal Standard and some Legal Rule, this implies that there would not be a misalignment per se, since there is no semblance of alignment inherently involved. Overall, for the lawful facet to be sustainable the Legal Rules and Legal Standards need to maintain some form of stasis.

### 3.2.2 No Legal Rules: No Legal Standards – *Lawlessness (a vacuum)*

In this use case, when there are No Legal Rules and No Legal Standards, the presumption is that this will lead to a state of being lawless since there is a vacuum as to what the legal rules are and what the legal standards are. Behavior is apparently allowed to free-range and does so without any guidance as to what is considered legally abiding. In this instance, the legal micro-directives would consist of an empty set.

### 3.2.3 Legal Rules: No Legal Standards – *Infinite Conundrum*

In this use case, when there are Legal Rules and No Legal Standards, the presumption is that this is potentially viable though raises the question of what the legal rules are predicated upon. In theory, overarching legal standards are the cornerstone for the formulation and cohesion of legal rules, without which the legal rules might seem arbitrary and seemingly absent of any overall pattern. In addition, legal rules without any corresponding legal standards can produce the so-called infinite conundrum. Namely, the number of legal rules might become massive, doing so to a degree that they potentially seem random and haphazard, lacking a structure or basis, and could stretch seemingly endlessly. The resultant legal micro-directives would potentially be perceived as a never-ending set that has little or no rationale or underlying constituency.

### 3.2.4 Legal Standards: No Legal Rules – *Vagueness Quandary*

In this use case, when there are Legal Standards and No Legal Rules, the presumption is that this is potentially viable though raises the difficulty of vagueness about appropriate and inappropriate behaviors. Legal standards provide an overarching semblance of legal behavior and yet allow for leeway and flexibility, but within that latitude also rests the possibility of inappropriate behaviors and thus without definitive legal rules to provide guidance it is potentially the matter that behaviors will flaunt standards and egress into what is ultimately ascertained as a lawless activity.

## 3.3 Macroscopic Process Flow of Legal Micro-Directives

As shown in **Figure B-3** and **Figure B-4**, it is useful to consider a macroscopic process flow that might underly the advent of legal micro-directives.

In Figure B-3, starting with the assumption that Legal Micro-Directives will flow from Legal Goals, there is a topmost process involving the creation of new legal goals. There is a baseline of existing legal goals that are utilized in the effort of new legal goals formulation. Once a new legal goal has been proposed, a legal micro-directives generator is invoked. There is next a reconciling of the generated legal micro-directives with the existing set of legal micro-directives, and a looping back to the formulation of Legal Goals is undertaken to contend with any legal micro-directives unable to be reconciled. After the reconciling has been accomplished, the promulgation of the legal micro-directives can be performed.

Figure B-4 is similar to Figure B-3, though this variant indicates the use of Legal Standards and Legal Rules, rather than the alternative proposition of Legal Goals and the elimination of Legal Standards and Legal Rules.

In Figure B-4, starting with the assumption that Legal Rules will flow from Legal Standards, there is a topmost process involving the creation of new legal standards. There is a baseline of existing standards that are utilized in the effort of new standards formulation. Once a new legal standard has been proposed, a legal rules generator is invoked, along with producing the legal micro-directives that are being postulated. There must be a reconciling of the Legal Rules, such that a comparison is made with the existing set of Legal Rules, and as needed a loop back to the formulation of Legal Standards is undertaken. After the reconciling has been accomplished, the promulgation of the legal micro-directives can be performed.

This is a macroscopic process flow and at a high-level does not encompass the variety of additional checks-and-balances and other facets that would be utilized for a more detailed delineation of the processes and sub-processes involved.

### 3.4 G.R.A.P.H.S. AI Enablement of Legal Micro-Directives

As shown in **Figure B-5**, AI enablement of Legal Micro-Directives consists of at least the following key AILR activities, denoted with an acronym of G.R.A.P.H.S. for convenience of reference:

- **Generate Legal Micro-Directives**
- **Reconcile Legal Micro-Directives**
- **Approve Legal Micro-Directives**
- **Promulgate Legal Micro-Directives**
- **Host Legal Micro-Directives**
- **Suspend Legal Micro-Directives**

Each of these AI activities is increasingly capable at each of the respective AILR Levels of Autonomy as discussed in the prior section of this paper. Per the macroscopic process flow indicated in the prior subsection, the AILR would to some degree (per the respective LoA's) be able to generate legal micro-directives, be able to reconcile legal micro-directives, be able to approve legal micro-directives, be able to promulgate legal micro-directives, and in addition be a hosting component that would be able to serve out legal micro-directives on an authenticated basis (as covered in the next subsection), and be able to suspend legal micro-directives (when so needed). Since the legal micro-directives would be considered a somewhat sacrosanct societal law renderer, the AI involved in the G.R.A.P.H.S. would need to be of the highest order of security, integrity, consistency, resiliency, etc.

In essence, the AI is tantamount to being the law giver, as it were, though presumably the Legal Goals or Legal Standards are the cornerstone that is the basis for the laws; nonetheless any semblance of AI system instabilities such as AI system corruption, disruption, or other maladies or malfunctions would indubitably undermine the collective belief in the bona fide legality of the micro-directives and thus wholly undermine the legal underpinnings of the legal micro-directives structure and philosophical foundation.

### 3.5 Brittleness of Legal Micro-Directives And Potential Remedies

As shown in **Figure B-6**, various brittleness effects can undermine the utility of legal micro-directives. This necessitates a consideration of the brittleness that can appear, along with how to mitigate or remedy these aspects, doing so to attempt at achieving a more robust advent of legal micro-directives.

The brittleness effects are listed as:
- Ripple effect
- Amalgamation effect
- Off-guard effect
- Propagation effect
- Wariness effect
- Conflicts effect
- Spoofing effect

In brief, each is described as:

**Ripple effect:** *Changes in Legal Standards or Legal Goals can produce a cavalcade of Legal Rules or Legal Micro-Directives changes, creating a massive ripple effect or torrent that is overwhelming and confounding*

**Amalgamation effect:** *Existing Legal Rules or Legal Micro-Directives become embedded fabric in societal activities; new Legal Rules or Legal Micro-Directives are disruptive to abiding with and costly or arduous to inure*

**Off-guard effect:** *Caught off-guard by the unexpected appearance of new Legal Rules or Legal Micro-Directives, lack of notification and forewarning and potentially inappropriate notice to the nature of the lawful matters involved*

**Propagation effect:** *Receiving of new Legal Rules or Legal Micro-Directives might encounter propagation delays, those seeking to heed are not provided a timely awareness, meanwhile still operating under the presumption of the validness of the prior set*

**Wariness effect:** *Wariness toward new Legal Rules or Legal Micro-Directives if all Legal Rules or Legal Micro-Directives seem to be unstable and rapidly modified/revoked/suspended*

**Conflicts effect:** *Multiple Legal Rules or Legal Micro-Directives that intrinsically conflict with each other*

**Spoofing effect:** *Spoofing to make illegitimate Legal Rules or Legal Micro-Directives seem as official*

The overarching emphasis is that these and additional brittleness facets that could be identified are all potential limitations, constraints, or outright threats to the veracity of the deployment and acceptance of Legal Micro-Directives and could therefore weaken or entirely undermine the adoption of Legal Micro-Directives. A concerted effort will be need to ensure that the manner of automation used for the enactment and enablement of the Legal Micro-Directives is able to address and mitigate or remedy these challenges.

For example, potential means of remedying or mitigating these delineated adverse effects consist of:

- Ripple effect: ***Reconciliation***
- Amalgamation effect: ***Disentanglement***
- Off-guard effect: ***Notification***
- Propagation effect: ***Attestation***
- Wariness effect: ***Stabilization***
- Conflicts effect: ***Harmonization***
- Spoofing effect: ***Authentication***

Details underlying each of these effects and their respective proposed remedies are described in Eliot [16] [17] [21].

**4 Additional Considerations and Future Research**

As earlier indicated, research by various legal scholars has advocated that the law might inevitably be transformed into legal micro-directives consisting of legal rules that are derived from legal standards. These legal rules are envisioned as being propagated via automation for everyday use as readily accessible lawful legal directives throughout society.

This paper has examined and sought to extend the legal micro-directives theories in three crucial respects: (1) By indicating that legal micro-directives are likely to be AI-enabled and evolve over time in scope and velocity across the autonomous levels of AI Legal Reasoning, (2) By exploring the tradeoffs between legal standards and legal rules as the imprinters of the micro-directives, and (3) By illuminating a set of brittleness exposures that can undermine legal micro-directives and proffering potential mitigating remedies to seek greater robustness in the instantiation and promulgation of such AI-enabled lawful directives.

Future research is needed to explore in greater detail the manner and means by which AI-enablement will occur, along with the potential for adverse consequences, including as conveyed the possibility of brittleness that could undermine the efficacy of legal micro-directives.

Additional research on the proposed mitigations or remedies of the legal micro-directive brittleness is also needed. If legal micro-directives are to be productively adopted, the full gamut of legal, economic, societal, and technological ramifications need to be sufficiently examined.

**About the Author**

Dr. Lance Eliot is the Chief AI Scientist at Techbrium Inc. and a Stanford Fellow at Stanford University in the CodeX: Center for Legal Informatics. He previously was a professor at the University of Southern California (USC) where he headed a multi-disciplinary and pioneering AI research lab. Dr. Eliot is globally recognized for his expertise in AI and is the author of highly ranked AI books and columns.

**Figure A-1**

| Level | Descriptor | Examples | Automation | Status |
|---|---|---|---|---|
| | **AI & Law: Levels of Autonomy For AI Legal Reasoning (AILR)** | | | |
| 0 | No Automation | Manual, paper-based (no automation) | None | De Facto - In Use |
| 1 | Simple Assistance Automation | Word Processing, XLS, online legal docs, etc. | Legal Assist | Widely In Use |
| 2 | Advanced Assistance Automation | Query-style NLP, ML for case prediction, etc. | Legal Assist | Some In Use |
| 3 | Semi-Autonomous Automation | KBS & ML/DL for legal reasoning & analysis, etc. | Legal Assist | Primarily Prototypes & Research Based |
| 4 | AILR Domain Autonomous | Versed only in a specific legal domain | Legal Advisor (law fluent) | None As Yet |
| 5 | AILR Fully Autonomous | Versatile within and across all legal domains | Legal Advisor (law fluent) | None As Yet |
| 6 | AILR Superhuman Autonomous | Exceeds human-based legal reasoning | Supra Legal Advisor | Indeterminate |

*Figure 1: AI & Law - Autonomous Levels by Rows*     *Source Author: Dr. Lance B. Eliot*     V1.3

**Figure A-2**

| | Level 0 | Level 1 | Level 2 | Level 3 | Level 4 | Level 5 | Level 6 |
|---|---|---|---|---|---|---|---|
| **Descriptor** | No Automation | Simple Assistance Automation | Advanced Assistance Automation | Semi-Autonomous Automation | AILR Domain Autonomous | AILR Fully Autonomous | AILR Superhuman Autonomous |
| **Examples** | Manual, paper-based (no automation) | Word Processing, XLS, online legal docs, etc. | Query-style NLP, ML for case prediction, etc. | KBS & ML/DL for legal reasoning & analysis, etc. | Versed only in a specific legal domain | Versatile within and across all legal domains | Exceeds human-based legal reasoning |
| **Automation** | None | Legal Assist | Legal Assist | Legal Assist | Legal Advisor (law fluent) | Legal Advisor (law fluent) | Supra Legal Advisor |
| **Status** | De Facto – In Use | Widely In Use | Some In Use | Primarily Prototypes & Research-based | None As Yet | None As Yet | Indeterminate |

AI & Law: Levels of Autonomy For AI Legal Reasoning (AILR)

*Figure 2: AI & Law - Autonomous Levels by Columns*  *Source Author: Dr. Lance B. Eliot*  V1.3

**Figure B-1**

| | Level 0 | Level 1 | Level 2 | Level 3 | Level 4 | Level 5 | Level 6 |
|---|---|---|---|---|---|---|---|
| | **Legal Micro-Directives: Levels of Autonomy For AI Legal Reasoning (AILR)** | | | | | | |
| **Descriptor** | No Automation | Simple Assistance Automation | Advanced Assistance Automation | Semi-Autonomous Automation | AILR Domain Autonomous | AILR Fully Autonomous | AILR Superhuman Autonomous |
| **Examples** | Manual, paper-based (no automation) | Word Processing, XLS, online legal docs, etc. | Query-style NLP, ML for case prediction, etc. | KBS & ML/DL for legal reasoning & analysis, etc. | Versed only in a specific legal domain | Versatile within and across all legal domains | Exceeds human-based legal reasoning |
| **Automation** | None | Legal Assist | Legal Assist | Legal Assist | Legal Advisor (law fluent) | Legal Advisor (law fluent) | Supra Legal Advisor |
| **Status** | De Facto – In Use | Widely In Use | Some In Use | Primarily Prototypes & Research-based | None As Yet | None As Yet | Indeterminate |
| **AI-Enabled Legal Micro-Directives** | n/a | Impractical | Incubatory | Infancy | Narrow | Wide | Consummate |

*Figure 1: Legal Micro-Directives - Autonomous Levels of AILR by Columns*     *Source Author: Dr. Lance B. Eliot*

V1.3

**Figure B-2**

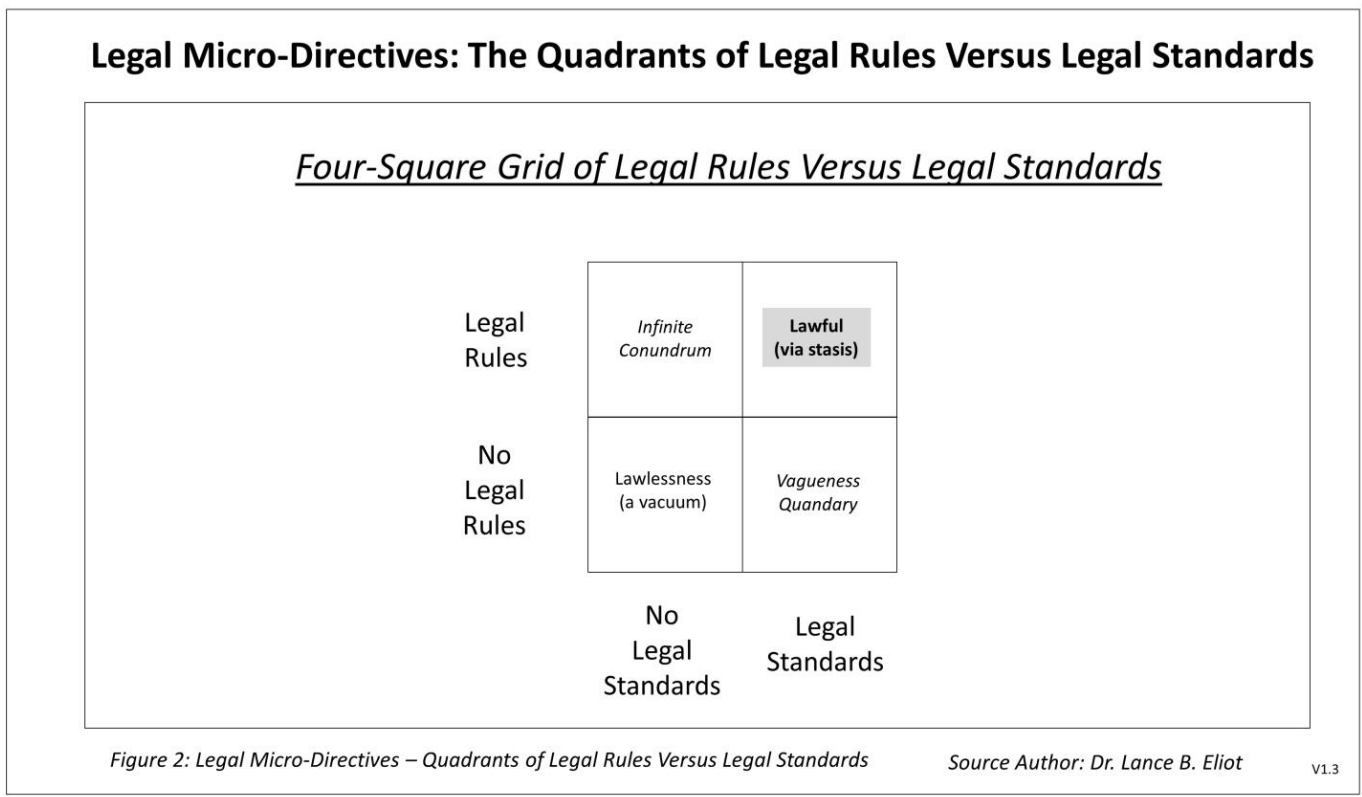

**Figure B-3**

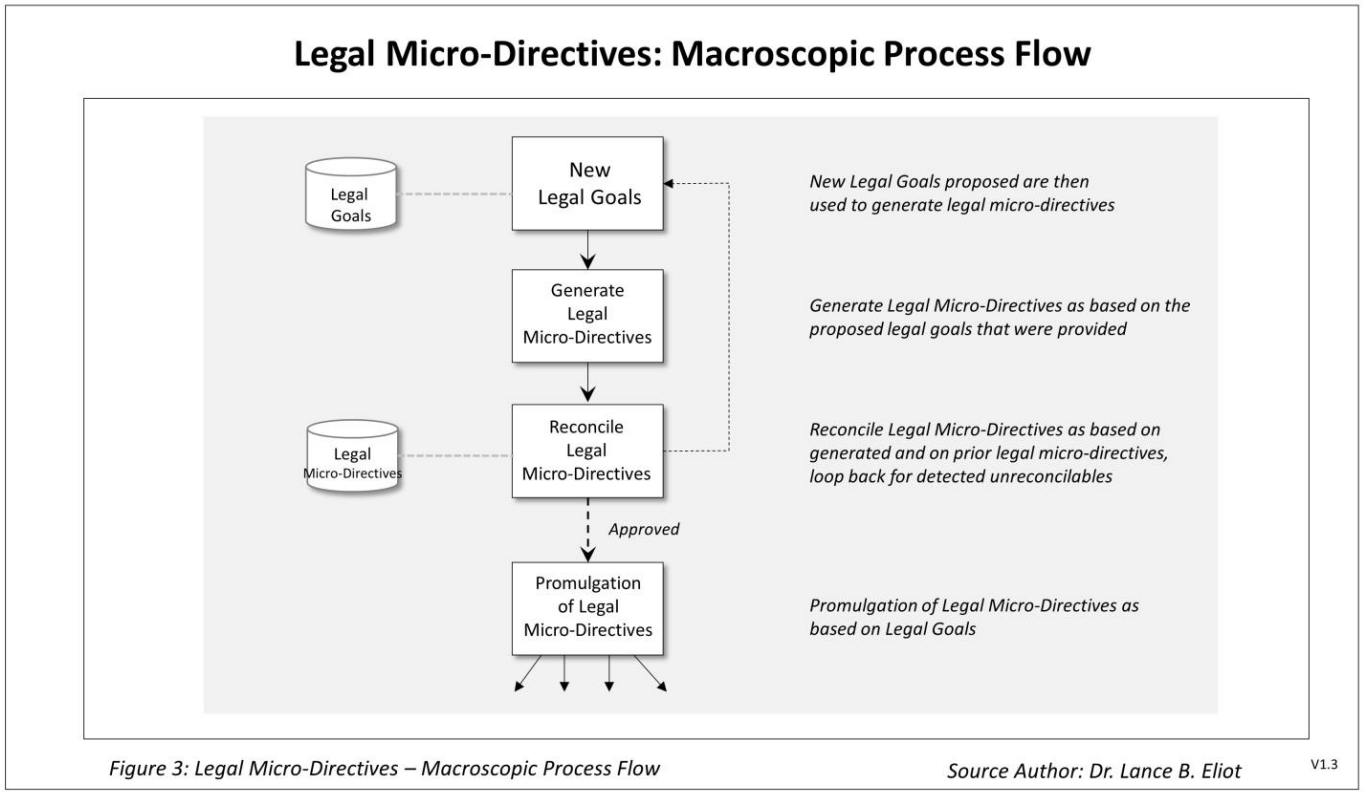

Figure 3: Legal Micro-Directives – Macroscopic Process Flow

Source Author: Dr. Lance B. Eliot

**Figure B-4**

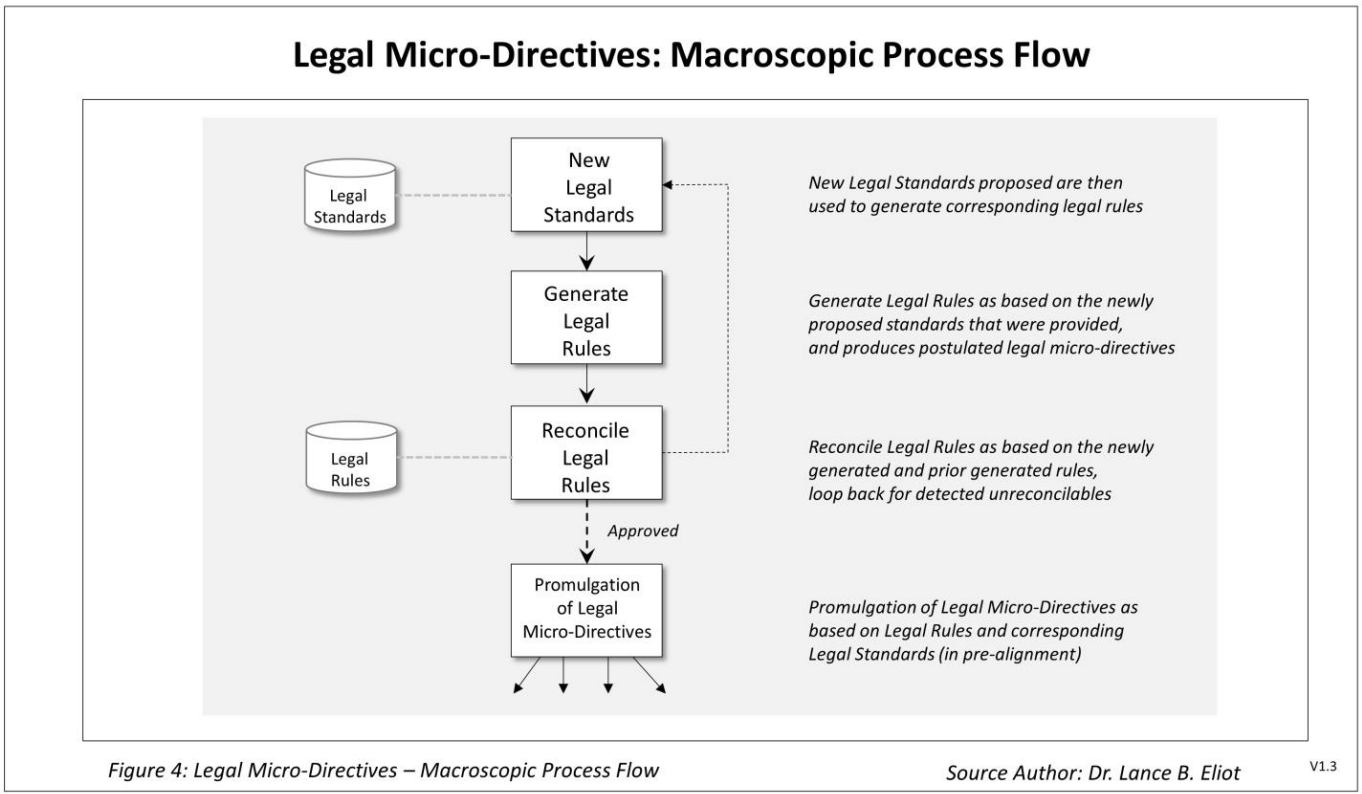

Figure 4: Legal Micro-Directives – Macroscopic Process Flow

**Figure B-5**

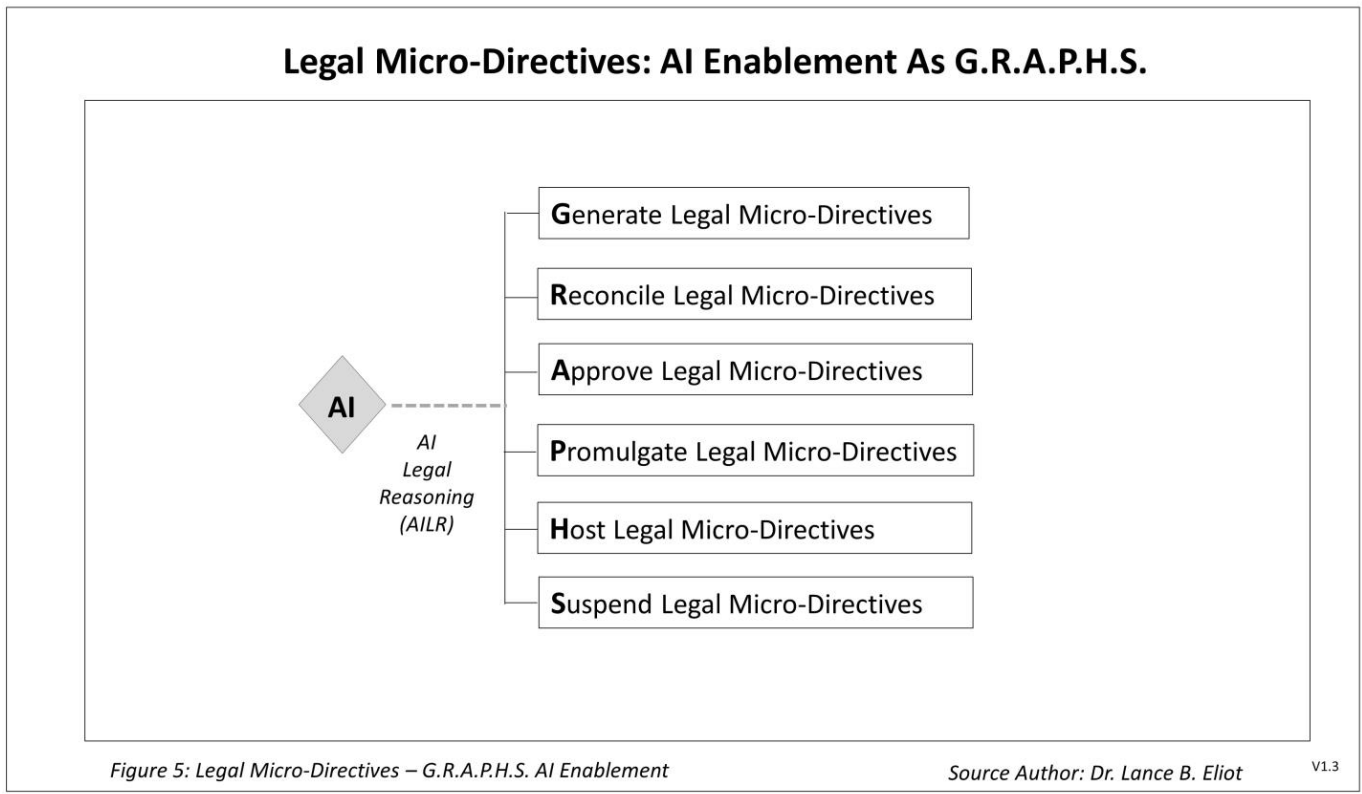

*Figure 5: Legal Micro-Directives – G.R.A.P.H.S. AI Enablement*   *Source Author: Dr. Lance B. Eliot*   V1.3

**Figure B-6**

## Legal Micro-Directives: Brittleness Aspects And Remedies

| Brittleness | Description | Potential Remedy |
|---|---|---|
| Ripple effect | Change in Standards can produce cavalcade of rules changes | Reconciliation |
| Amalgamation effect | Existing rules become fabric; new rules disruptive to abide | Disentanglement |
| Off-guard effect | Caught off-guard by unexpected appearance of new rules | Notification |
| Propagation effect | Receiving of new rules might encounter propagation delays | Attestation |
| Wariness effect | Wariness toward new rules if all rules seem to be unstable | Stabilization |
| Conflicts effect | Multiple rules that intrinsically conflict with each other | Harmonization |
| Spoofing effect | Spoofing to make illegitimate rules seem as official | Authentication |

*Figure 6: Legal Micro-Directives – Brittleness and Remedies*　　　*Source Author: Dr. Lance B. Eliot*　V1.3